# Harnessing orbital Hall effect for energy-efficient magnetization switching in room-temperature van der Waals ferromagnet $Fe_3GaTe_2$

*Chenhui Zhang, Hua Bai, Yuchen Pu, and Hyunsoo Yang**

Department of Electrical and Computer Engineering, National University of Singapore, Singapore 117583, Singapore

E-mail: eleyang@nus.edu.sg

**Abstract**

2D van der Waals (vdW) magnets provide new opportunities for spin–orbit torque magnetoresistive random-access memory (SOT-MRAM) due to their unique properties. Electrically manipulating the magnetization of vdW magnets is key to realizing 2D SOT-MRAM, whereas conventional spin Hall materials such as heavy metals and topological insulators suffer from limitations in torque efficiency and energy consumption. Although recent studies show that the orbital Hall conductivity in light metals greatly exceeds the spin Hall conductivity, direct experimental demonstrations that the orbital Hall effect (OHE) can induce more energy-efficient SOT switching than the spin Hall effect in vdW magnets remain scarce. Here, we utilize Cr as the orbital current source to efficiently manipulate the magnetization of the vdW ferromagnet $Fe_3GaTe_2$ at room temperature. In the $Fe_3GaTe_2$/Pt (1.5 nm)/Cr (4.5 nm) trilayer structure, the orbital current originating from Cr is converted into the spin current via Pt, which then exerts a torque on $Fe_3GaTe_2$. Compared with control samples using 6 nm Pt as the spin current source, the switching current density in OHE-based devices are reduced by ~3.9 times, resulting in a ~52% reduction in power consumption. This work presents the promising potential of harnessing orbital currents to realize energy-efficient 2D SOT-MRAM.

## 1. Introduction

The rapid rise of emerging technologies, particularly artificial intelligence (AI), is imposing ever-increasing demands on memory devices. Spin–orbit torque magnetoresistive random-access memory (SOT-MRAM) offers a promising pathway toward high-speed, high-endurance, and energy-efficient nonvolatile data storage technology. In SOT-MRAM, the spin current originating from the spin source layer is injected into the ferromagnetic layer and exerts a torque on the magnetic moments to switch the magnetization (i.e. to write the bits). Therefore, both the spin source and ferromagnetic layers play important roles in determining device performance. Unlike conventional ferromagnetic materials such as CoFeB, emerging 2D magnets may provide new opportunities for SOT-MRAM.[1] The long-range magnetic order in 2D ferromagnets, such as $Cr_2Ge_2Te_6$,[2] $Fe_3GeTe_2$,[3] and $Fe_3GaTe_2$[4] can be sustained even down to atomic thickness while maintaining perpendicular magnetic anisotropy (PMA), a key property for high-density spintronic devices. More importantly, the atomically smooth surfaces of 2D magnetic materials can provide an excellent interface between the spin source and ferromagnetic layers, leading to a high spin transparency.[5] In addition, the room-temperature 2D ferromagnet $Fe_3GaTe_2$ exhibits a much larger coercive field than CoFeB,[4] which enhances magnetic immunity against external perturbations, such as stray fields and Oersted fields, thereby reducing the risk of unintended switching. The higher anisotropy associated with large coercivity improves thermal stability, suppressing thermally induced magnetization fluctuations. As a result, $Fe_3GaTe_2$ is promising for scaling down device dimensions while maintaining reliable operation, which is advantageous for achieving higher-density magnetic memory.

For spin current source materials, heavy metals such as Pt, Ta, and W are commonly used due to their strong spin–orbit coupling (SOC), which converts charge currents into spin currents via the spin Hall effect (SHE) and interfacial Rashba effect.[6-8] The SOT switching of 2D ferromagnets is first demonstrated in the $Fe_3GeTe_2$/Pt system,[9, 10] which stimulates a surge of research interest in 2D SOT-MRAM. Alternatively, topological insulators and semimetals can also serve as spin current sources through spin-polarized surface states.[11-18] By integrating topological materials with the high-Curie-temperature ($T_C$) 2D ferromagnet $Fe_3GaTe_2$, current-induced magnetization switching is achieved in Sn-doped $Bi_{1.1}Sb_{0.9}Te_2S/Fe_3GaTe_2$,[19] $WTe_2/Fe_3GaTe_2$,[20] and $TaIrTe_4/Fe_3GaTe_2$[21, 22] van der Waals (vdW) heterostructures at room temperature. However, conventional spin Hall materials face limitations in torque efficiency and energy consumption. Heavy metals are typically good conductors, but their torque efficiencies are relatively low.[23] In contrast, topological materials generally exhibit higher torque efficiencies but lower electrical conductivities, leading to significant current shunting and power dissipation issues.[24]

Recently, the orbital Hall effect (OHE) has attracted much attention, as it can generate an orbital current analogous to the SHE.[25-27] The OHE originates from the momentum-space orbital texture, which is ubiquitous and independent of the strength of SOC.[27] Therefore, a wide range of materials can be utilized as orbital current sources, such as light metals Ti,[28, 29] Cr,[30, 31] Zr,[32, 33] Mn,[34] and Ru.[35] More importantly, the intrinsic orbital Hall conductivity (OHC) in light metals greatly exceeds the spin Hall conductivity (SHC),[36] potentially leading to enhanced torque efficiency and reduced energy consumption. To date, orbital-current-induced magnetization switching has been demonstrated in various material systems, including conventional magnetic thin films and 2D magnets. [33-35, 37-40] Despite these advances, important questions remain to be addressed. Previous studies have revealed that in light metals, the orbital relaxation length can reach several tens of nanometers, and the torque efficiency continues to increase within this range.[28, 41] Therefore, thick orbital source layers are usually employed to achieve enhanced torque efficiency and reduced switching current density.[33, 34, 40] However, increasing the thickness inevitably raises the total switching current. In comparison, the widely used spin Hall material Pt reaches a saturated SOT efficiency at a thickness of only ~6 nm.[42] Thus, it remains unclear whether orbital Hall materials still offer significant advantages in energy consumption compared with 6 nm Pt when used at the same thickness.

In this work, we harness the OHE to achieve efficient room-temperature magnetization switching of 2D vdW ferromagnet $Fe_3GaTe_2$. The light metal Cr is employed as the orbital current source. With the assistance of a thin Pt interlayer, the orbital current originating from Cr is converted into the spin current and exerts a torque on $Fe_3GaTe_2$. Unlike previous studies that typically employed thick orbital source layers, we set the Cr and Pt thicknesses to 4.5 nm and 1.5 nm, respectively. In the control samples,  the prototypical 6 nm Pt spin Hall material is used. With the same thickness of SOT sources, the total switching current is nearly proportional to the switching current density in different samples, which enables a fair comparison of the switching performance. Current-induced magnetization switching experiments reveal that the switching current density and switching efficiency in OHE-based devices are reduced and enhanced by approximately 3.9 and 4.7 times, respectively, leading to a ~52% reduction in power consumption. Our results highlight the potential of OHE-based 2D SOT-MRAM for energy-efficient data storage.

## 2. Results and Discussion

### 2.1. Device architecture for efficient orbital-to-spin conversion

To realize high-efficiency orbital-current-induced magnetization switching, an effective orbital current source is essential. Theoretical calculations predict that Cr is a good orbital Hall material possessing an OHC as large as $\sigma_{\mathrm{OH}}^{\mathrm{Cr}}$= 5829 ($\hbar/e$) ($\Omega$ cm)$^{-1}$, while its SHC is negligibly small and shows an opposite sign, i.e. $\sigma_{\mathrm{SH}}^{\mathrm{Cr}}$= −162 ($\hbar/e$) ($\Omega$ cm)$^{-1}$.[43] Normally, the local magnetic moment in the ferromagnetic layer cannot directly receive a torque from the orbital current, since it has no exchange coupling with the orbital angular momentum (**L**). Thus, a thin conversion layer with strong SOC is needed to convert the orbital angular momentum into spin angular momentum (**S**).[31] Based on the above considerations, the device is designed as a trilayer structure, as illustrated in Figure 1a. When a charge current is injected into the device, the orbital current generated by Cr (i.e. $J_L^{Cr}$) flows into the Pt interlayer because of the OHE. Since Pt has strong and positive SOC, corresponding to a positive orbital-to-spin conversion coefficient $\eta_{L\text{–}S}$, the orbital current $J_L^{Cr}$ is converted into a spin current $J_{L\text{–}S}$ as it passes through Pt. Another spin current $J_S^{Cr}$ generated by SHE in Cr also flows into $Fe_3GaTe_2$ but has an opposite sign, as denoted by the blue arrow in Figure 1a. As the SHC of Cr is significantly smaller than the OHC and it further decays in the Pt layer, $J_S^{Cr}$ should be negligibly small. Moreover, the Pt layer can also contribute a spin current $J_S^{Pt}$, which has a same sign with $J_{L\text{–}S}$. Here, the Pt thickness is set to 1.5 nm to maximize the torque efficiency by considering the tradeoff between the **L–S** conversion and the decay of $J_{L\text{–}S}$ across the Pt layer.[44] To enable a fair comparison with the commonly used 6 nm-thick Pt SOT source, the Cr layer thickness is set to 4.5 nm, bringing the total thickness of Pt/Cr to 6 nm. In this case, the switching current density $J_c$ is used to evaluate the switching performance, since the total switching current $I_c$ is nearly proportional to $J_c$ when the device thicknesses are same.

The $Fe_3GaTe_2$ single crystals used in this study are grown by chemical vapor transport (CVT) method.[45] The characterization results for the chemical composition and crystal structure are presented in Figure S1 and S2, respectively. The $T_C$ of bulk crystals is ~ 366 K (see Figure S3). Figure 1b shows optical images of an exfoliated $Fe_3GaTe_2$ flake after sputtering with Pt (1.5 nm)/Cr (4.5 nm)/$SiO_2$ (4 nm)/$TaO_x$ (1.5 nm) on top (left) and the corresponding Hall bar device (right), where $SiO_2$ (4 nm)/$TaO_x$ (1.5 nm) serves as the capping layer (see fabrication details in the Experimental Section). The thickness of $Fe_3GaTe_2$ in the device area is 16.2 nm, which is determined by atomic force microscopy (AFM). The Hall resistance loops in Figure 1c reveal that the PMA of the flake device can be sustained at room temperature. The temperature-dependent coercive field $\mu_0 H_c$ and anomalous Hall resistance $R_{xy}^{AH}$ are extracted in Figure 1d and 1e, respectively. According to the $R_{xy}^{AH}$ vs. temperature data in Figure 1e, the $T_C$ of the device is estimated to be between 340 and 350 K, which is slightly lower than that of the bulk sample, but almost the same as in pristine flakes (see Figure S4e), indicating that the Pt/Cr layer does not significantly alter the magnetic properties of $Fe_3GaTe_2$.

**2.2. Orbital-current-induced magnetization switching of $Fe_3GaTe_2$**

Next, we demonstrate the OHE-driven magnetization switching in $Fe_3GaTe_2$/Pt/Cr devices. For clarity, the devices composed of $Fe_3GaTe_2$/Pt (1.5 nm)/Cr (4.5 nm) and $Fe_3GaTe_2$/Pt (6 nm) are hereafter denoted as PC# and P#, respectively. Figure 2a and 2b display the room-temperature Hall resistance loops of device PC#1 under an external magnetic field swept along the out-of-plane ($H||c$) and in-plane ($H||ab$) directions, respectively. A coercive field of $\mu_0 H_c = 14.6$ mT and a magnetic anisotropy field of $\mu_0 H_k = 2.8$ T are extracted, indicating the robust PMA at room temperature. In the current-induced magnetization switching experiments, a DC current pulse with a duration of 100 μs is applied to switch the magnetization, followed by a 500 μA read pulse applied after 3 s to measure the Hall resistance $R_{xy}$. The switching current density is determined using the effective current flowing through the SOT source layer after accounting for current shunting (Supplementary Note 1). As shown in Figure 2c, the room-temperature current-induced magnetization switching of $Fe_3GaTe_2$ is successfully achieved under an in-plane magnetic field of $\mu_0 H_x = 50$ mT along the current direction. Note that $m_z$ is calculated as the ratio of measured Hall resistance to the anomalous Hall resistance, i.e. $R_{xy}/R_{xy}^{AH}$. The switching ratio is around 56%, which is close to the ones in $Fe_3GeTe_2$/Pt and $Fe_3GaTe_2$/Pt bilayer devices in previous studies.[9, 46, 47] The incomplete switching can be attributed to several reasons, such as multi-domain state caused by Joule heating, domain-wall pinning, and interfacial damage during the device fabrication.[48, 49]

When the external field direction is flipped, the polarity of the switching loop is also reversed (Figure 2d), which is consistent with the conventional SOT switching behavior. The critical switching current density $J_c$ is determined to be $3.84 \times 10^6$ A cm$^{-2}$ at $\mu_0 H_x = 50$ mT, which is significantly smaller than $1.3 \times 10^7$ A cm$^{-2}$ in a recent study using 6 nm Pt as the spin current source.[47] Figure 2e exhibits the field dependence of switching behavior. Since the in-plane field is required to break the symmetry, no deterministic switching is observed at $\mu_0 H_x = 0$ T. The critical switching current density remains nearly constant under different assist fields, in contrast to conventional CoFeB-based SOT devices, where $J_c$ typically decreases with increasing assist field.[50] Similar phenomenon is also observed in $Fe_3GaTe_2$/Pt bilayer devices in previous studies,[46, 47, 51] which can be attributed to the significantly larger magnetic anisotropy field in $Fe_3GaTe_2$.

Polar magneto-optical Kerr effect (PMOKE) microscopy is employed to visualize the magnetization switching. The device investigated in the PMOKE experiments, shown in Figure S5b (device PC#2), consists of $Fe_3GaTe_2$ (13.8 nm)/Pt (1.5 nm)/Cr (4.5 nm)/$SiO_2$ (4 nm)/$TaO_x$ (1.5 nm). First, we perform the current-induced switching by measuring Hall resistance. As shown in Figure S5d,e, deterministic switching is achieved under an assist field of ± 50 mT at 300 K. The $J_c$ is $4.19 \times 10^6$ A cm$^{-2}$, comparable to that of PC#1. In the PMOKE experiments, the device is kept under ambient conditions at ~294 K. With an assist field of 70 mT along the $+x$ direction, we initialize the magnetization state of $Fe_3GaTe_2$ to the $+z$ direction by applying a current pulse of 5 mA ($J_c = 6.75 \times 10^6$ A cm$^{-2}$), and subtract the background. Afterwards, a current pulse of 5 mA is injected along the $-x$ direction. One can observe that the contrast of the $Fe_3GaTe_2$ device turns white, consistent with the magnetization switching from the $+z$ to $-z$ direction, as shown in Figure 3a. Subsequently, we subtract the background again and apply a current pulse of 5 mA along the $+x$ direction, which makes the contrast turn black, indicating the down to up magnetization switching. The magnetic contrast change is reversed when the assist field is applied to the $-x$ direction, as shown in Figure 3b. In the meantime, some unswitched regions are also observed, which can be attributed to the nonuniform domain nucleation and domain-wall propagation during the current-induced switching process. In exfoliated $Fe_3GaTe_2$ flakes, local pinning centers, such as structural defects, thickness inhomogeneity, edge roughness, and residual strain introduced during device fabrication, may hinder domain-wall motion and prevent complete magnetization reversal in certain areas. Similar incomplete switching behaviors are also reported in other vdW ferromagnetic systems.[48, 49] Nevertheless, the overall PMOKE and anomalous Hall measurements (Figure S5) consistently demonstrate deterministic current-induced magnetization switching in our devices.

**2.3. Comparison of spin- and orbital-current-induced magnetization switching**

To compare the switching performance of $Fe_3GaTe_2$-based devices using orbital Hall materials versus conventional spin Hall materials, control experiments are conducted with a Pt single layer serving as the SOT source. As illustrated in the top panel of Figure 4a, the thickness of the Pt single layer is fixed at 6 nm, which allows a fair comparison with the Pt (1.5 nm)/Cr (4.5 nm) bilayer source. The optical image of P#1 is shown in Figure S7a, which consists of $Fe_3GaTe_2$ (11.3 nm)/Pt (6 nm)/$SiO_2$ (4 nm)/$TaO_x$ (1.5 nm). First, we examine its magnetic properties. According to the temperature-dependent $R_{xy}^{\mathrm{AH}}$ data (Figure S7d), the $T_C$ of P#1 lies between 340 and 350 K, which is the same as that of PC#1 in Figure 1e. As shown in Figure S8a,b, the coercive field and magnetic anisotropy field at 300 K are 12.4 mT and 2.8 T, respectively, which are also comparable to those of PC#1. As shown in Figure S8c–e, the $J_c$ of P#1 is $1.64 \times 10^7$ A cm$^{-2}$ under $\mu_0 H_x = 50$ mT at 300 K (Figure S8c), which is close to previous studies.[47] However, this value is considerably larger than that of PC#1, revealing the superiority of orbital Hall materials over conventional spin Hall materials in switching 2D ferromagnet $Fe_3GaTe_2$.

Furthermore, current-induced magnetization switching experiments are performed at various temperatures ranging from 310 to 250 K on both devices P#1 and PC#1, as displayed in Figure 4a and 4b, respectively. At all measured temperatures, the $J_c$ of PC#1 is smaller than that of P#1, as summarized in Figure 4c. For instance, the $J_c$ is $2.84 \times 10^6$ A cm$^{-2}$ for PC#1 and $1.31 \times 10^7$ A cm$^{-2}$ for P#1 at 310 K, which increase to $6.76 \times 10^6$ A cm$^{-2}$ and $2.80 \times 10^7$ A cm$^{-2}$, respectively, at 250 K. The increased $J_c$ at lower temperatures can be attributed to the enhanced PMA and saturation magnetization of $Fe_3GaTe_2$. The great reduction of $J_c$ in PC#1 indicates the promoted switching efficiency enabled by the Pt/Cr bilayer SOT source.

To verify reproducibility, multiple devices are measured (Figure S6, S9 and S10), and the results are summarized in Table S1. Despite slight variations in the $Fe_3GaTe_2$ thickness, each device group exhibits similar switching performance, revealing the reproducibility of the results. The average $J_c$ in Pt (1.5 nm)/Cr (4.5 nm) bilayer and Pt (6 nm) single-layer SOT sources are $4.25 \times 10^6$ A cm$^{-2}$ and $1.65 \times 10^7$ A cm$^{-2}$, respectively, at 300 K. Considering the micrometer-scale lateral dimensions of the devices, the magnetization switching occurs via domain nucleation and domain-wall propagation.[52, 53] In this regime, the switching efficiency can be evaluated from the current-induced switching results through $\xi = \frac{4e}{\pi \hbar}\mu_0 M_s t_{FM}\left(\frac{H_c}{J_c}\right)$,[54] where $e$ is the electron charge, $\hbar$ is the reduced Planck constant, $\mu_0$ is the vacuum permeability, and $M_s$ and $t_{FM}$ represent the saturation magnetization and thickness of the ferromagnet layer, respectively. As shown in Figure 4d, for the $Fe_3GaTe_2$/Pt (6 nm) devices, the average $\xi$ at 300 K is calculated to be 0.46, within the range reported for conventional Pt-based SOT systems (i.e. 0.23–0.48).[55] On the other hand, the average $\xi$ in $Fe_3GaTe_2$/Pt (1.5 nm)/Cr (4.5 nm) devices reaches 2.17, ~ 4.7 times that of the $Fe_3GaTe_2$/Pt (6 nm) control samples. This value is even comparable to those reported in topological-material-based 2D ferromagnet SOT devices, e.g. $\xi$ = 1.49 in $TaIrTe_4/Fe_3GaTe_2$ at 300 K[21] and $\xi$ = 4.6 in $WTe_2/Fe_3GeTe_2$ at 150 K.[56]

Further, we perform harmonic Hall measurements to evaluate the torque efficiency.[57, 58] The estimated damping-like torque efficiencies are 0.10 and 0.35 for the Pt (6 nm) and Pt (1.5 nm)/Cr (4.5 nm) SOT sources, respectively (Supplementary Note 2 and Figure S11). The torque efficiency is therefore enhanced by a factor of ~ 3.5 in the Pt (1.5 nm)/Cr (4.5 nm) structure. This result is consistent with the significantly reduced switching current density observed in the switching measurements. For Pt, it is known that its torque efficiency increases with increasing the thickness before reaching saturation (~6 nm).[59] In Supplementary Note 3, we demonstrate that the critical switching current densities are nearly identical in the $Fe_3GaTe_2$/Pt (6 nm) and $Fe_3GaTe_2$/Pt (7 nm) devices (Figure S12). Moreover, the spin current generated by the 1.5 nm Pt alone is insufficient to switch the magnetization of $Fe_3GaTe_2$ (Figure S13). These results indicate that the large torque efficiency observed in the Pt (1.5 nm)/Cr (4.5 nm) devices cannot be attributed primarily to the thin Pt layer, but instead mainly originates from the additional contribution associated with the Cr layer. In addition, we perform magnetization switching measurements on $Fe_3GaTe_2$/Cr (6 nm) bilayer devices (Supplementary Note 4). No deterministic switching is observed until device failure occurs due to the large applied current (Figure S14). Therefore, the observed large torque efficiency should be primarily attributed to the strong orbital Hall effect in Cr and the effective orbital-to-spin conversion in Pt, which collectively give rise to a large $J_{L-S}$.

Finally, we evaluate the energy consumption for spin- and orbital-current-induced magnetization switching in $Fe_3GaTe_2$. The switching power consumption of the entire device can be calculated as $P = \rho J^2$, in which $\rho$ and $J$ are the resistivity and current density of the device, respectively. They can be obtained by $\rho = \frac{\rho_{FM}\rho_{source}(t_{FM}+t_{source})}{\rho_{FM}t_{source}+\rho_{source}t_{FM}}$ and $J = \frac{I_c}{w(t_{FM}+t_{source})}$, where $w$ is the channel width, $I_c$ is the critical switching current, and $\rho_{FM}$ ($\rho_{source}$) and $t_{FM}$ ($t_{source}$) are the resistivity and thickness of the ferromagnet (spin source) layer, respectively. As shown in Figure 4d, the average switching power consumption for the $Fe_3GaTe_2$/Pt (6 nm) and $Fe_3GaTe_2$/Pt (1.5 nm)/Cr (4.5 nm) devices are $3.43 \times 10^{12}$ mW cm$^{-3}$ and $1.66 \times 10^{12}$ mW cm$^{-3}$, respectively. Notably, the results reveal that the power consumption can be reduced by ~52% in orbital-current-induced switching devices. Unlike previous studies,[40] which generally employ relatively thick orbital source layers to enhance the torque efficiency, our work demonstrates enhanced torque efficiency and reduced switching current density under thickness-matched conditions, enabling a fair comparison with conventional Pt-based SOT devices. More importantly, we demonstrate a substantial reduction of switching power consumption for $Fe_3GaTe_2$ at room temperature, highlighting the potential of orbital current engineering for energy-efficient 2D spintronic applications.

## 3. Conclusion

We have demonstrated energy-efficient OHE-driven magnetization switching of 2D ferromagnet $Fe_3GaTe_2$ at room temperature. The SOT device is designed as an $Fe_3GaTe_2$/Pt (1.5 nm)/Cr (4.5 nm) trilayer structure, where the light metal Cr serves as the orbital current source and Pt is responsible for the orbital-to-spin conversion. In the current-induced magnetization switching experiments, both Hall resistance measurements and PMOKE imaging are performed to verify the SOT switching of perpendicular magnetization in $Fe_3GaTe_2$. An average critical switching current density of $4.25 \times 10^6$ A $cm^{-2}$ is obtained from multiple devices, which is ~ 3.9 times smaller than that of the $Fe_3GaTe_2$/Pt (6 nm) SHE-based control samples. The reduced switching current density is attributed to the enhanced switching efficiency originating from the OHE. Finally, a ~52% reduction in power consumption is achieved in OHE-based devices. Our results demonstrate the promising potential of OHE for realizing energy-efficient 2D SOT-MRAM.

## 4. Experimental Section

*Crystal growth and characterizations*: The $Fe_3GaTe_2$ single crystals were grown using the CVT method.[45] Fe (99.95%, Aladdin), GaTe (99.99%, Macklin), and Te (99.999%, Alfa Aesar) powders with a mole ratio of 3:1:1 were used as the starting materials. They were additionally mixed with a small amount of iodine (transport agent) and loaded into quartz ampoules. Afterwards, the quartz ampoules were evacuated and sealed. After one-week growth in a tube furnace with a temperature gradient of 750–700 °C, the ampoules were naturally cooled to room temperature. The chemical composition of the crystals was determined by energy-dispersive X-ray spectroscopy on a Hitachi Regulus 8230 scanning electron microscope. The X-ray diffraction patterns were acquired using a Rigaku SmartLab diffractometer. The magnetization measurements were performed in a physical property measurement system with the VSM option (DynaCool, Quantum Design).

*Device fabrication and electrical transport measurements*: The $Fe_3GaTe_2$ flakes were mechanically exfoliated onto a $Si/SiO_2$ (300 nm) substrate using blue tapes (Ultron Systems). Then the samples were immediately loaded into a magnetron sputtering system with a base pressure of less than $5 \times 10^{-9}$ Torr. After sputtering the Pt (6 nm) or Pt (1.5 nm)/Cr (4.5 nm) SOT source layer, all the samples were further deposited with a $SiO_2$ (4.0 nm)/Ta (1.5 nm) capping layer. The Ta layer naturally oxidized into $TaO_x$ after being exposed to air upon removal from the chamber. The flake thickness was determined by atomic force microscopy (Park NX20). In the first lithography step, the Hall bars were patterned by electron-beam lithography (Raith eLINE) using ma-N 2403 negative tone photoresist and etched by Ar-ion milling (Scia Coat 200). In the second lithography step, the electrodes were patterned using ultraviolet maskless lithography (Tuotuo Technology), followed by sputtering deposition of Ta (5 nm)/Cu (80 nm)/Pt (10 nm). The electrical transport measurements were performed in a physical property measurement system (EverCool, Quantum Design). In the current-induced magnetization switching experiments, a DC current pulse with a duration of 100 μs was applied by a Keithley 6221 current source. After 3 s, a 500 μA read pulse with a duration of 10 ms was applied to measure the Hall resistance using a Keithley 2182A nanovoltmeter. In the PMOKE (TuoTuo Technology) experiments, the devices were kept under ambient conditions at ~294 K. In the harmonic Hall measurements, the Keithley 6221 was used to generate the alternating current with a frequency of 37.1 Hz. The first and second harmonic Hall voltage signals were collected by lock-in amplifiers (Stanford SR830).

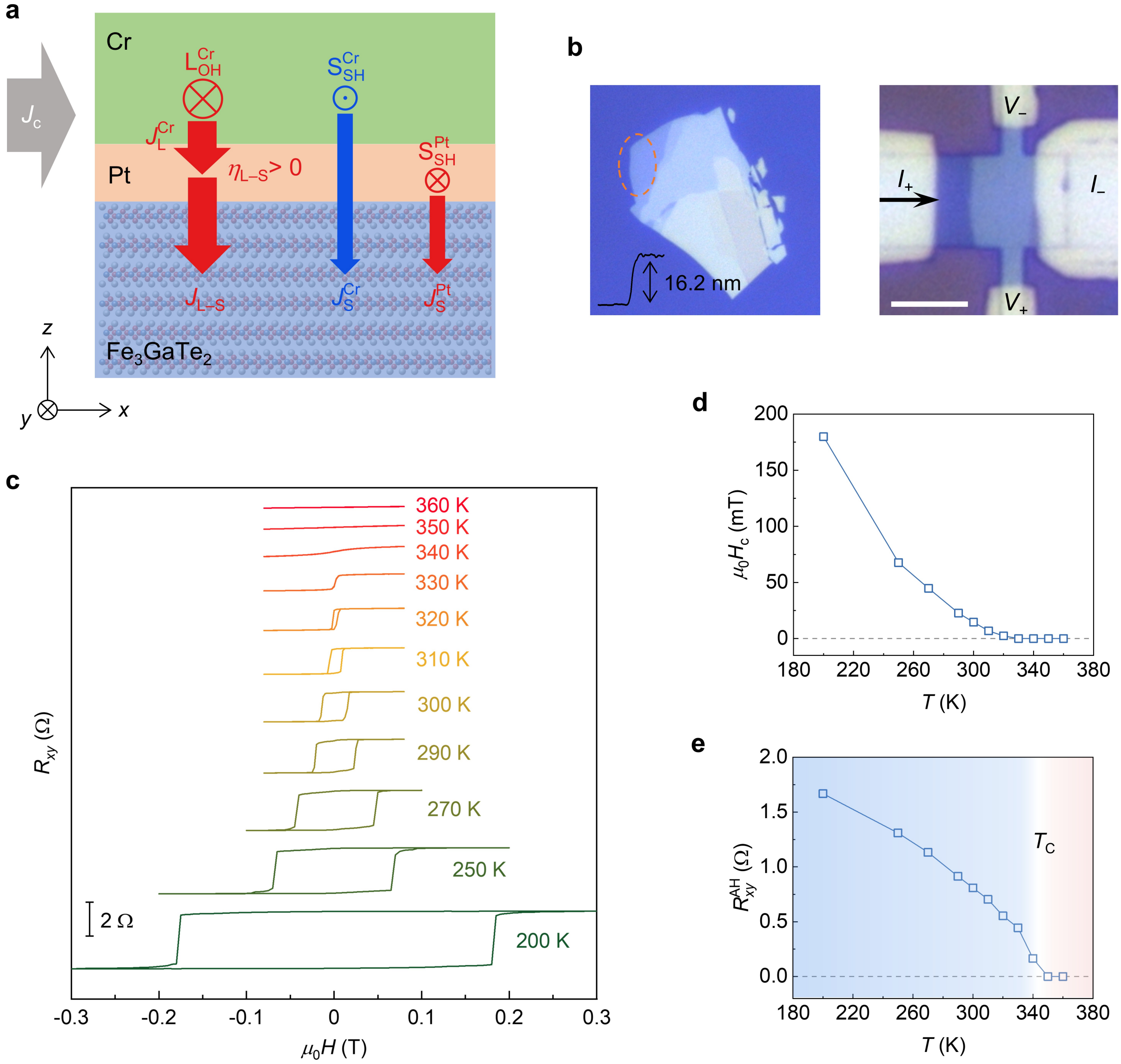

**Figure 1.** Basic properties of the $Fe_3GaTe_2$/Pt/Cr device. a) Schematic illustration of spin and orbital currents generation and conversion in the $Fe_3GaTe_2$/Pt/Cr trilayer. b) Optical images of an exfoliated $Fe_3GaTe_2$ flake after sputtering Pt (1.5 nm)/Cr (4.5 nm)/$SiO_2$ (4.0 nm)/$TaO_x$ (1.5 nm) on top (left) and the corresponding Hall bar device (right) (device PC#1). The black step curve in the inset is the AFM height profile of $Fe_3GaTe_2$ in the device area (outlined by the orange dashed circle). The scale bar is 5 μm. c) Hall resistance loops measured at various temperatures. d) Temperature-dependent coercive field $\mu_0H_c$ extracted from (c). e) Temperature-dependent anomalous Hall resistance $R_{xy}^{AH}$.

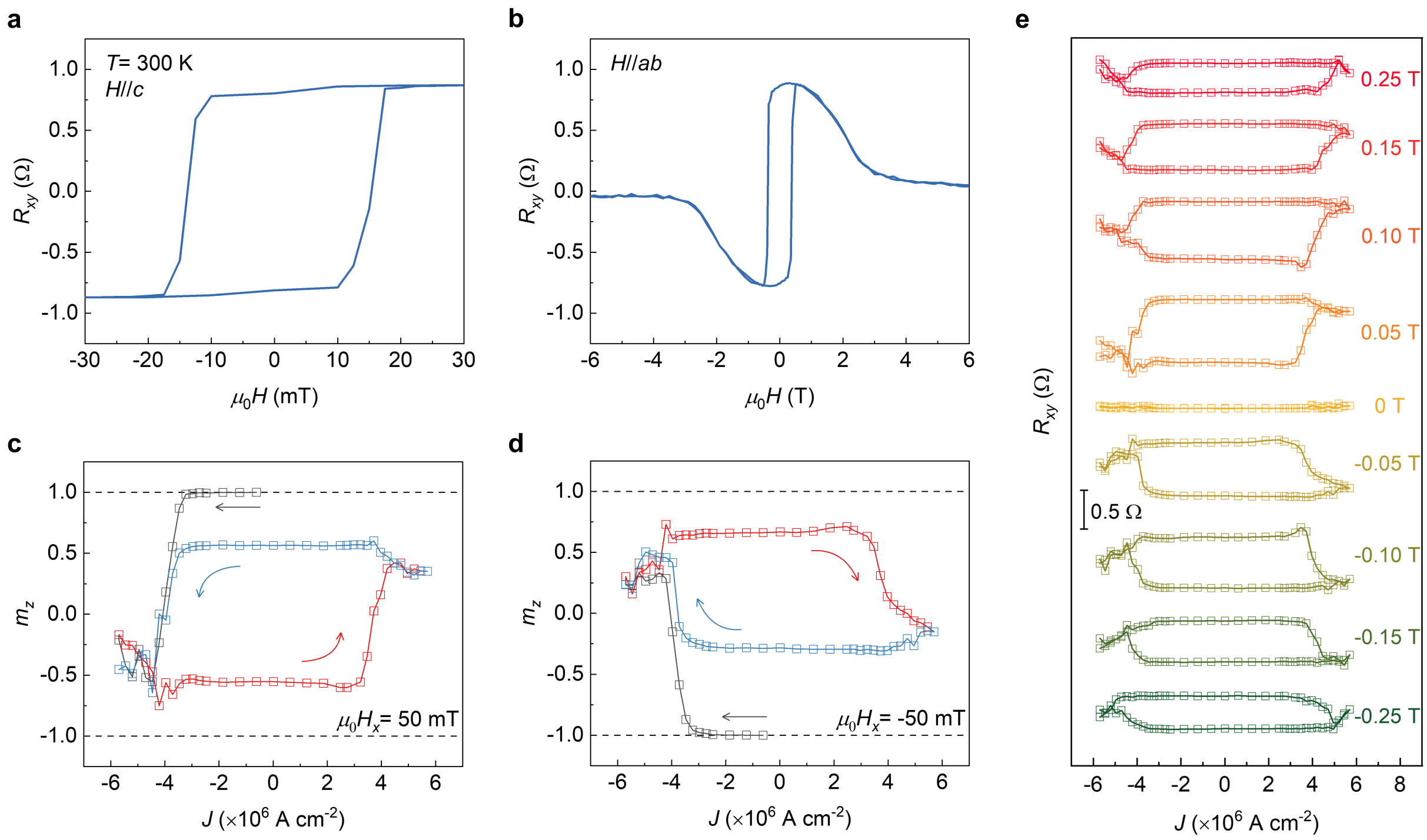

**Figure 2.** Current-induced magnetization switching in $Fe_3GaTe_2$ (16.2 nm)/Pt (1.5 nm)/Cr (4.5 nm) (device PC#1). a,b) Hall resistance loops measured under out-of-plane (a) and in-plane (b) magnetic fields at 300 K. c,d) Room-temperature current-induced magnetization switching measured under $\mu_0 H_x = 50$ mT (c) and −50 mT (d). $m_z$ is calculated as the ratio of measured Hall resistance to the anomalous Hall resistance. e) Magnetization switching loops measured under different in-plane magnetic fields at 300 K.

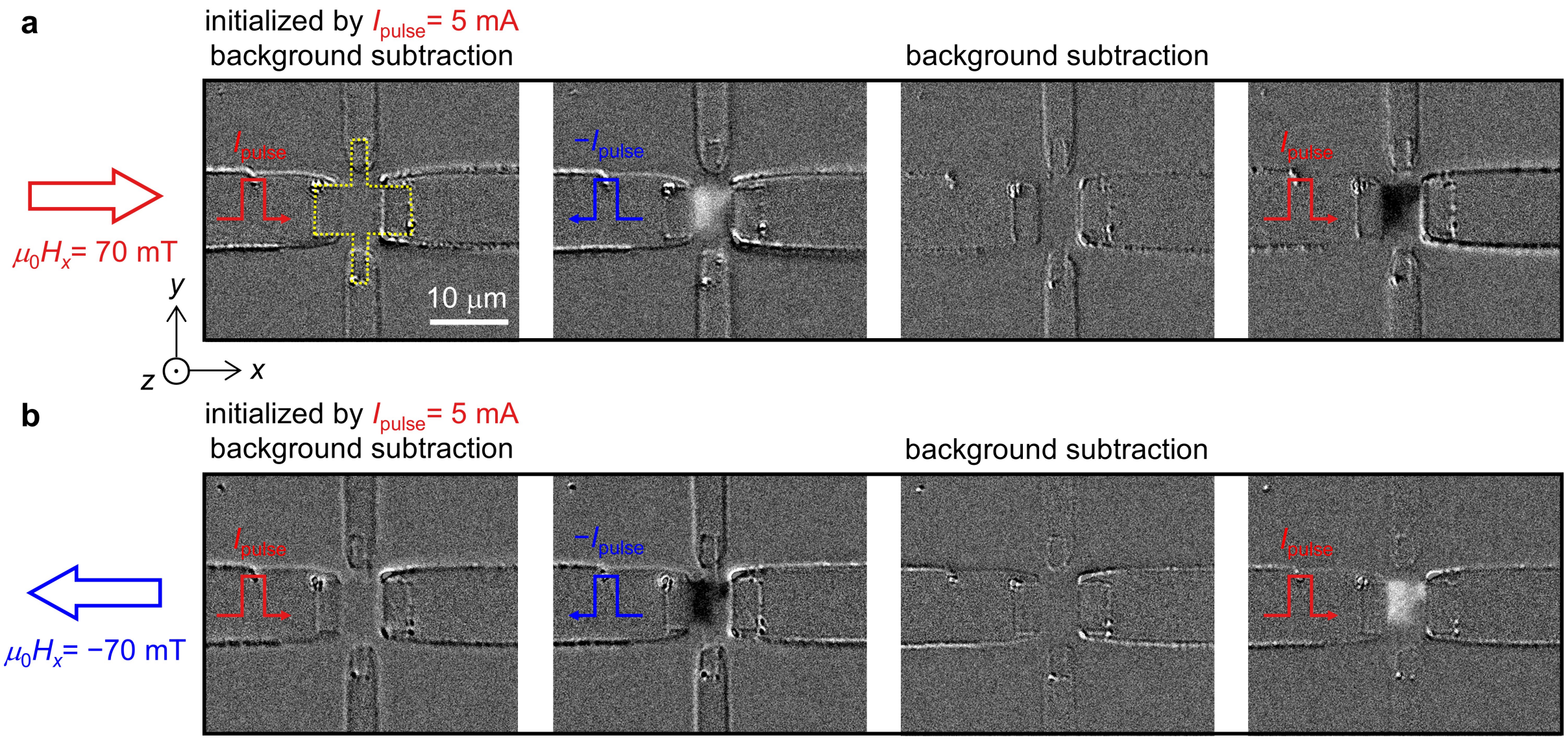

**Figure 3.** PMOKE images for current-induced magnetization switching in $Fe_3GaTe_2$ (13.8 nm)/Pt (1.5 nm)/Cr (4.5 nm) (device PC#2). a,b) An external magnetic field of 70 mT is applied along the $+x$ (a) and $-x$ (b) direction. A current pulse of 5 mA with a duration of 100 μs is applied to switch the magnetization. The device area is outlined by yellow dashed lines. The white and black contrasts indicate the magnetization pointing along $-z$ and $+z$ directions, respectively. The measurements are conducted under ambient conditions at ~294 K.

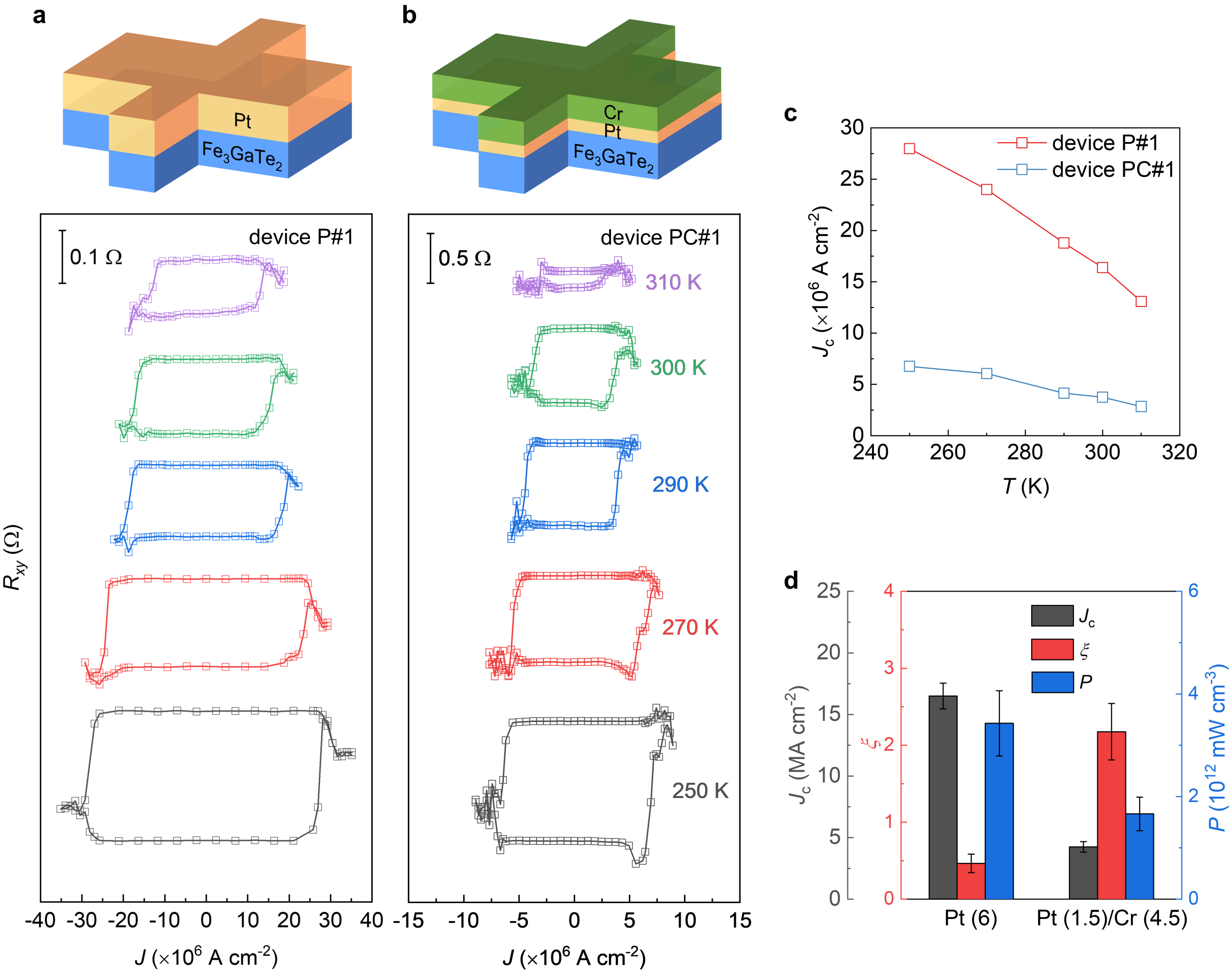

**Figure 4.** Comparison of spin- and orbital-current-induced magnetization switching. a,b) Device structures and magnetization switching loops of (a) $Fe_3GaTe_2$ (11.3 nm)/Pt (6 nm) (device P#1) and (b) $Fe_3GaTe_2$ (16.2 nm)/Pt (1.5 nm)/Cr (4.5 nm) (device PC#1). c) Temperature-dependent critical switching current densities. d) Comparison of average critical switching current density $J_c$, switching efficiency $\xi$, and switching power consumption $P$ among multiple devices that use different SOT sources.